,

# Extending Field Limits in Nanoscale Magnetic Imaging with Metamaterial-inspired Magnetic Flux Concentrators

A. Barrera[1,‡], E. Fourneau[2,‡], T. Pirottin[2,3], L. Marcano[4], R. Abrudan[5], R. Huang[6], Ll. Balcells[1], I. Orue[7], M.L. Fdez-Gubieda[8], D. Villanueva[8], A.G. Gubieda[9], L. Chang[6], B. Vanderheyden[3], R.J. Harrison[10], A.V. Silhanek[2,*], A. Palau[1,*] and S. Valencia[5,*]

A. Barrera, Ll. Balcells and Anna Palau
[1]Institut de Ciència de Materials de Barcelona, ICMAB-CSIC, Campus de la UAB, 08193 Bellaterra, Spain

Emile Fourneau, Thomas Pirottin and Alejandro V. Silhanek
[2]Experimental Physics of Nanostructured Materials, Q-MAT, Department of Physics, Université de Liège, Sart-Tilman B-4000, Belgium

Thomas Pirottin and Benoît Vanderheyden
[3]Montefiore Research Unit, Department of Electrical Engineering and Computer Science, Université de Liège, B-4000 Sart Tilman, Belgium

Lourdes Marcano
[4]Departamento de Física, Universidad de Oviedo, Calvo Sotelo s/n, 33007, Oviedo, Spain

Radu Abrudan and Sergio Valencia
[5]Helmholtz-Zentrum Berlin fur Materialien und Energie GmbH, Berlin, Germany

Rong Huang and Liao Chang
[6]Peking University, Beijing, China

Iñaki Orue
[7]SGIker, Universidad del País Vasco – UPV/EHU, 48940 Leioa, Spain

María Luisa Fdez-Gubieda and Danny Villanueva
[8]Dpto. Electricidad y Electrónica, Universidad del País Vasco, 48940 Leioa, Spain

Alicia G. Gubieda
[9]Dpto. Inmunología, Microbiología y Parasitología, Universidad del País Vasco – UPV/EHU, 48940 Leioa, Spain

Richard J. Harrison
[10]Department of Earth Sciences, University of Cambridge, U.K.

[‡]Equal contributing authors

*Corresponding authors: sergio.valencia@helmholtz-berlin.de, palau@icmab.es, and asilhanek@uliege.be

,

**Abstract**

Many nanoscale magnetic imaging techniques are constrained by the maximum magnetic field that can be applied during measurements, due to geometrical limitations or interactions with the probe or the detected signal (e.g., electrons). Here, it is demonstrated that sample-integrated metamaterial-inspired magnetic flux concentrators (MFCs) locally amplify magnetic fields, allowing observation of magnetization processes beyond instrumental limits. Micrometer-sized MFCs fabricated directly on the samples are tested in photoemission electron microscopy experiments employing X-ray magnetic circular dichroism as magnetic contrast mechanism. At low applied fields, substantial amplification factors enable observation of magnetization reversal in a chain of magnetite nanoparticles synthesized by magnetotactic bacteria at an applied field of 8 mT, substantially smaller than the ~50 mT predicted by simulations in absence of MFCs. At higher fields, the field enhancement extends the accessible field range by a factor of five, enabling for the first time, imaging of the field-dependent magnetic domain structure evolution of an isolated giant magnetofossil. Finally, we show how MFC geometry and material parameters can be tuned to optimize performance considering sample and experimental constraints, providing a tunable and broadly applicable strategy for extending the accessible field range in a wide variety of nanoscale magnetic imaging techniques.

## 1. Introduction

Progress in materials science is closely linked to the development of novel experimental capabilities able to address relevant open questions. In particular, research areas such as, spintronics[1-3], topological magnetism[4,5], magnetization dynamics[6], multiferroics[7] and nanomagnetism[8] demand spatially resolved and magnetically sensitive techniques capable of visualizing spin textures and magnetic domains as a function of an externally applied magnetic field[9,10]. Nanoscale magnetic imaging, therefore, stands as a cornerstone in modern condensed matter physics[11], providing unique insights into magnetic coupling[3], magnetic anisotropy[12], exchange and dipolar interactions[13], pinning and depinning of magnetic domain walls[14], stability of magnetic topological structures[15] and magnetic phase transitions[16,17]. Unfortunately, many of these techniques are inherently limited by the strength of magnetic fields that can be applied during imaging, either due to spatial constraints[18] or the interaction of the magnetic field with the detected signal[19] or the probe[20]. For example, electron-based microscopy techniques such as scanning electron microscopy with polarization analysis, electron holography, Lorentz transmission electron microscopy, spin-polarized low-energy electron microscopy and photoemission electron microscopy (PEEM) are particularly affected because the Lorentz force deflects electron trajectories on their way to the detector. This effect, which can partially

,

be compensated by microscopy optics, ultimately leads to blurring, distortion and reduced spatial resolution of the resulting magnetic image.

Here we present a novel approach to overcome this fundamental limitation by employing metamaterial-inspired, sample-integrated magnetic flux concentrators (MFCs). Although the term metamaterial originally referred to periodic arrays of subwavelength resonant elements, its use has expanded in recent years to include artificially structured materials designed to manipulate low-frequency and even static magnetic fields through magnetostatic transformation optics[21,22]. In this regime, magnetic and electric fields decouple, and tailored field responses can be achieved using non-resonant structures whose effective permeability arises from their engineered geometry rather than from the intrinsic properties of the constituent material[23].

Within this framework, transformation optics prescribes that the maximum in magnetic field concentration is achieved in the limit where the radial permeability $\mu_r \rightarrow \infty$ and the azimuthal one $\mu_\theta \rightarrow 0$ (Ref. [22]). Flower-shaped MFCs (Figure 1a) provide one such realization, where the geometry itself produces the required anisotropic permeability tensor ($\mu_r \gg \mu_\theta$) and enables efficient guidance and concentration of magnetic flux toward the core. Although the overall structure is radially symmetric, the presence of the gap explicitly breaks the angular isotropy, enabling efficient channeling of magnetic flux into the gap along the direction of the applied field. Within this context, Barrera et al. demonstrated the scalability of flower-shaped MFCs down to the nanoscale, achieving a two-order-of-magnitude enhancement in magnetic sensor sensitivity by concentrating the magnetic field at their center[24].

Building on this concept, we explore the integration of flower-shaped MFCs directly into the sample architecture to locally concentrate the externally applied magnetic field at the region of interest, thereby enabling effective field strengths beyond those currently achievable.

Expanding the accessible magnetic field range during magnetic imaging would significantly broaden the number of materials and nanoscale systems that could be characterized. These include systems with field- and temperature-dependent magnetic phase transitions[17], exchange bias systems[25], artificial spin-ice[13,26], magnetic nanoparticles and nanostructures[14], and antiferromagnetic spintronic devices such as spin valves[27] and tunnel magnetoresistance junctions[28], including 2D van der Waals magnets[17,29].

We demonstrate the viability of this approach using photoemission electron microscopy with X-ray magnetic circular dichroism (XMCD-PEEM) as magnetic contrast mechanism[30]. Among magnetic imaging techniques, XMCD-PEEM stands out due to its surface-sensitivity, high spatial resolution,

,

element-selectivity, and ability to visualize buried layers and/or interfaces[31]. Moreover, control over the polarization of the incoming photon beam allows the characterization of ferro-, ferri-, and anti-ferromagnetic compounds[32]. Custom-designed sample holders (Figure 1a), confining stray magnetic fields closer to the sample surface, enable magnetic fields up to about ±30 mT to be applied during imaging[19]. Nevertheless, this field range restricts in-field characterization to soft ferromagnetic systems. So far, semi-hard and hard ferromagnetic systems have remained inaccessible, requiring indirect approaches to explore their field dependent properties, such as applying a high magnetic field pulse and measuring at magnetic remanence[5,13,14], limiting the capability to explore their field-dependent properties directly.

Sample-integrated flower-shaped MFCs overcome this limitation. Their functionality is assessed using two paradigmatic systems i) a magnetosome chain, consisting of high-quality 45 nm single-domain magnetite nanoparticles naturally synthesized by magnetotactic bacteria[33] and measured at a temperature below their Verwey transition, and ii) a spearhead-shaped magnetite giant magnetofossil[34] ca. 2 µm long with a 1 µm base diameter. These systems, spanning distinct dimensional regimes, and with saturation fields significantly larger than 30 mT, provide an ideal platform to evaluate the feasibility of the proposed strategy across both nanoscale and microscale magnetic structures.

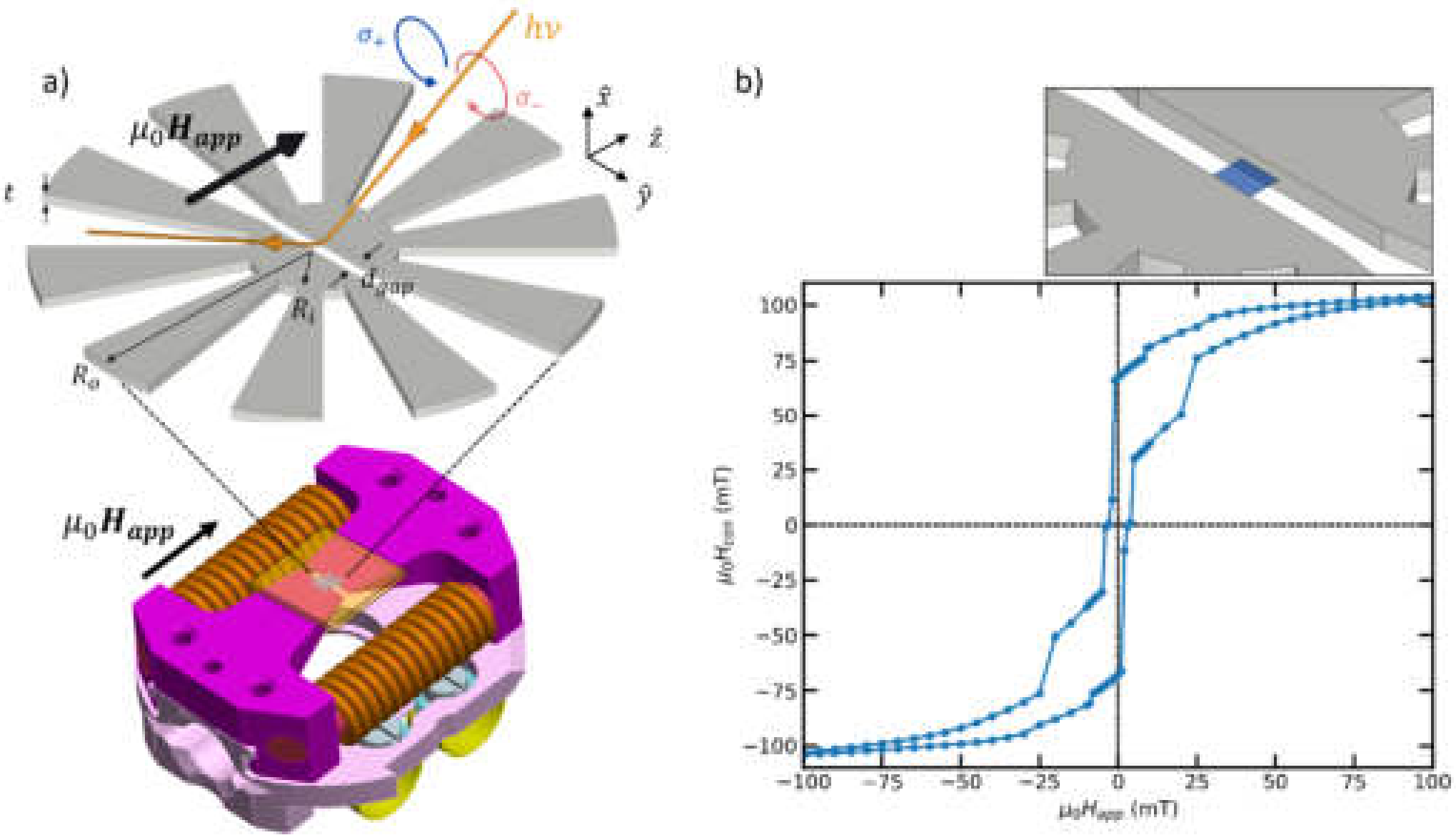

**Figure 1** a) Schematic of a custom-made PEEM sample holder with a sample-integrated magnetic flux concentrator. The magnetic field ($\mu_0 H_{app}$) applied during imaging is generated by current-driven coils (orange) mounted on the holder and guided by a magnetic yoke (purple) towards the center of the pole gap. With this holder, maximum magnetic fields during imaging of up to ±30 mT can be reached (ref. [19]) without MFC. The MFC, deposited on top of a substrate (shown transparent for clarity), is positioned within this gap. An enlarged view of the MFC is shown above the sample holder, along with the geometrical parameters characterizing the MFC. b) Micromagnetic simulations. The graph shows the average magnetic field

,

created by the MFC in the blue-colored spot as a function of the applied in-plane magnetic field. See the main text for exact dimensions of the MFC used for the simulation.

## 2. Results and discussion

### 2.1 Sample description and magnetic flux concentrator integrated XMCD-PEEM geometry

Magnetosome chains and giant magnetofossils samples were deposited onto Si substrates. Scanning electron microscopy was employed to identify isolated chains and fossils with minimal surrounding magnetic material introduced during the deposition process, thereby reducing potential undesired magnetic interactions with the MFCs. Cobalt magnetic flux concentrators were subsequently fabricated on selected samples, positioning the target structure at the center of their gap (see methods section).

XMCD images as a function of the applied magnetic field have been obtained at the Co- and Fe $L_3$-edges to characterize the magnetic response of both the MFC and the ferromagnetic sample being studied. The magnetic field is applied across the gap, perpendicular to gap edges. The incoming X-ray beam is incident at 16° relative to the sample surface, with its in-plane propagation direction aligned along the field direction, as indicated in Figure 1a. In this geometry the XMCD signal is proportional to the projection of the magnetization along the gap-crossing direction (z).

The effective field created by an MFCs at a given applied magnetic field $\mu_0 H_{app}$ can be expressed as

$$\mu_0 H_{eff} = \mu_0 H_{app} + \mu_0 H_{con} \quad \text{eqn (1)}$$

where $\mu_0 H_{con}$ is the magnetic field created by the concentrator. This contribution arises from the magnetization of the concentrator and therefore varies as a function of the applied field $\mu_0 H_{app}$. In this work, the ratio $\mu_0 H_{eff}/\mu_0 H_{app}$ is defined as the concentration gain to distinguish it from the intrinsic magnetic susceptibility or relative permeability of cobalt, which describes material properties rather than the geometry-induced flux guiding effects central to MFC operation.

For flower-shaped magnetic metasurfaces, $\mu_0 H_{con}$ at their gap center depends on the material employed, on geometrical factors such as the thickness ($t$), internal ($R_I$) and outer ($R_O$) radii, gap size ($d_{gap}$), number of petals[35] (Figure 1a), and on the local magnetic domain distribution [36] (see supporting information section). Figure 1b shows micromagnetic simulations for a cobalt MFC with 10 petals, $R_O = 5R_I = 10$ µm, thickness of 70 nm, and 500 nm gap. Step-like magnetization jumps observed in the micromagnetic simulations arise due to abrupt changes of magnetic domain arrangement such as the formation of flux -closure states. A more gradual magnetization reversal is expected for larger structures as they allow a larger number of domains and a more gradual and spatially averaged

,

magnetization reversal. An effective field of ca. 50 mT can theoretically be obtained with an applied field ($\mu_0 H_{app}$) of just 5 mT, which represents a gain $H_{eff}/H_{app}$ of 10. Comparable gain values may occur in simple circular Co patches at sufficiently high fields, where the response is mainly governed by magnetic saturation and the associated stray field near the gap. At lower fields, however, the flower-like geometry tends to provide higher gains due to its anisotropic flux-guiding behavior, a characteristic that simple geometries do not generally exhibit. In addition, increasing the overall magnetic volume of the concentrator tends to enhance the attainable effective field and the resulting gain, as a larger ferromagnetic volume contributes more strongly to the stray-field focusing within the gap. The influence of thickness and other key geometric parameters on the concentrator performance are modelled and discussed in detail in section 2.4 and Supporting Information.

Accurate micromagnetic simulations of such large-scale devices providing the relationship between $\mu_0 H_{app}$ and $\mu_0 H_{eff}$ become computationally unfeasible due to the required processing time and/or GPU memory. Alternatively, the field dependence of this relationship can, in principle, be determined experimentally in test MFCs using embedded magnetic sensors, such as by tracking the displacement of a vortex core in a ferromagnetic disk placed at the center of the MFC[37] or by measuring the planar Hall effect in a ferromagnetic reference element[35]. However, the results of such a direct measurement must be interpreted with caution, as they can yield a correspondence between $\mu_0 H_{app}$ and $\mu_0 H_{eff}$ which does not reflect the experimental reality. This discrepancy may arise from extrinsic factors such as, for example, structural imperfections or geometric variations between the test structure and the actual MFC used in the microscopy experiments. These differences, which can be introduced during fabrication (see for instance the partially closed gap in Figure 2a), can lead to noncomparable magnetic domain distributions caused by magnetic domain pinning. Moreover, the presence within the concentrator area of additional magnetic material resulting from preparation of the targeted magnetic structures to be investigated (see, Figures 2 and 3) locally distorts the magnetic flux.

To overcome this limitation the effective field as a function of $\mu_0 H_{app}$ might be obtained by imaging and measuring the field-dependent magnetic response of the employed MFC[36].

**2.2 Validation of magnetic flux concentrators using a magnetosome chain**

Magnetic nanoparticles synthesized by magnetotactic bacteria, known as magnetosomes, have gained significant relevance in biomedical applications and magnetic materials research owing to their genetically controlled properties[38]. When assembled into linear chains, they form a one-dimensional natural magnetic assembly, providing an ideal playground to study fundamental magnetic properties and enabling direct comparison between experiments and micromagnetic simulations. In this work, magnetosome chains are used as a stable and well-understood benchmark system for validating the

,

performance of sample-integrated magnetic flux concentrators and for demonstrating that XMCD-PEEM can access magnetic-field regimes beyond its current limit.

A flower-shaped cobalt magnetic flux concentrator with a thickness $t$ = 70 nm, outer radius $R_o$=100 µm, inner radius $R_i$ = 20 µm, and a nominal gap width $d_{gap}$ = 500 nm was fabricated surrounding a ca. 500 nm long chain composed of 10 magnetosomes synthesized by *Magnetospirillum gryphiswaldense* MSR-1 strain (Figure 2). Magnetite magnetosomes produced by this strain have an average diameter of 45 nm (Ref. [39]).

The fabrication process resulted in a chain positioned within the gap of the concentrator (Figure 2b) but slightly misoriented from the direction across to the gap (~20°), which defines the direction of the in-plane magnetic field.

The magnetic domain state of the system was imaged by XMCD-PEEM (see methods section). Field-dependent data were acquired at T = 50 K. Below the Verwey transition temperature of magnetite (~120 K), the interplay of uniaxial magnetocrystalline anisotropy, shape anisotropy, and dipolar coupling between neighboring particles results in a strong temperature-dependent increase of the coercive field, reaching approximately 50 mT[40] at T = 50 K, which exceeds current in-field PEEM imaging capabilities.

Figures 2c and 2d show X-ray absorption spectroscopy (XAS) images for the concentrator and the magnetosome chain obtained at the Co $L_3$- and Fe $L_3$-edges, respectively. Even though the size of the magnetosomes is close to the instrumental resolution limit precluding visualization of the individual nanoparticles, the chain is well distinguished (Figures 2d). This is also evidenced in the corresponding XMCD images (Figs. 2e-f) showing that the magnetic domain state of the chain can be well resolved and separated from that of the magnetosome cluster to its right.

Spatially mapped and element-selective magnetic hysteresis loops were obtained by acquiring XMCD images as a function of the nominally applied magnetic field ($\mu_0 H_{app}$). At each field, the XMCD signal is integrated over the chain and over an MFC region close to it (Figure 2e) to obtain the hysteresis loops, depicted in Figures 2g-h. Both loops present hysteresis. For the MFC, the coercive field is ca. $\mu_0 H_c$ ~ 5 mT, and its saturation field $\mu_0 H_s$ is approximately 8 mT. The loop corresponding to the magnetosome chain is slightly wider, with $\mu_0 H_c$ ~ $\mu_0 H_s$ ~ 7-8 mT. The latter are approximately a factor 7 times smaller than those predicted by simulations ($\mu_0 H_c$ ~50 mT), where the magnetosome chain is considered as a collection of independent magnetic dipoles[39], see Figure 2i (solid line).

,

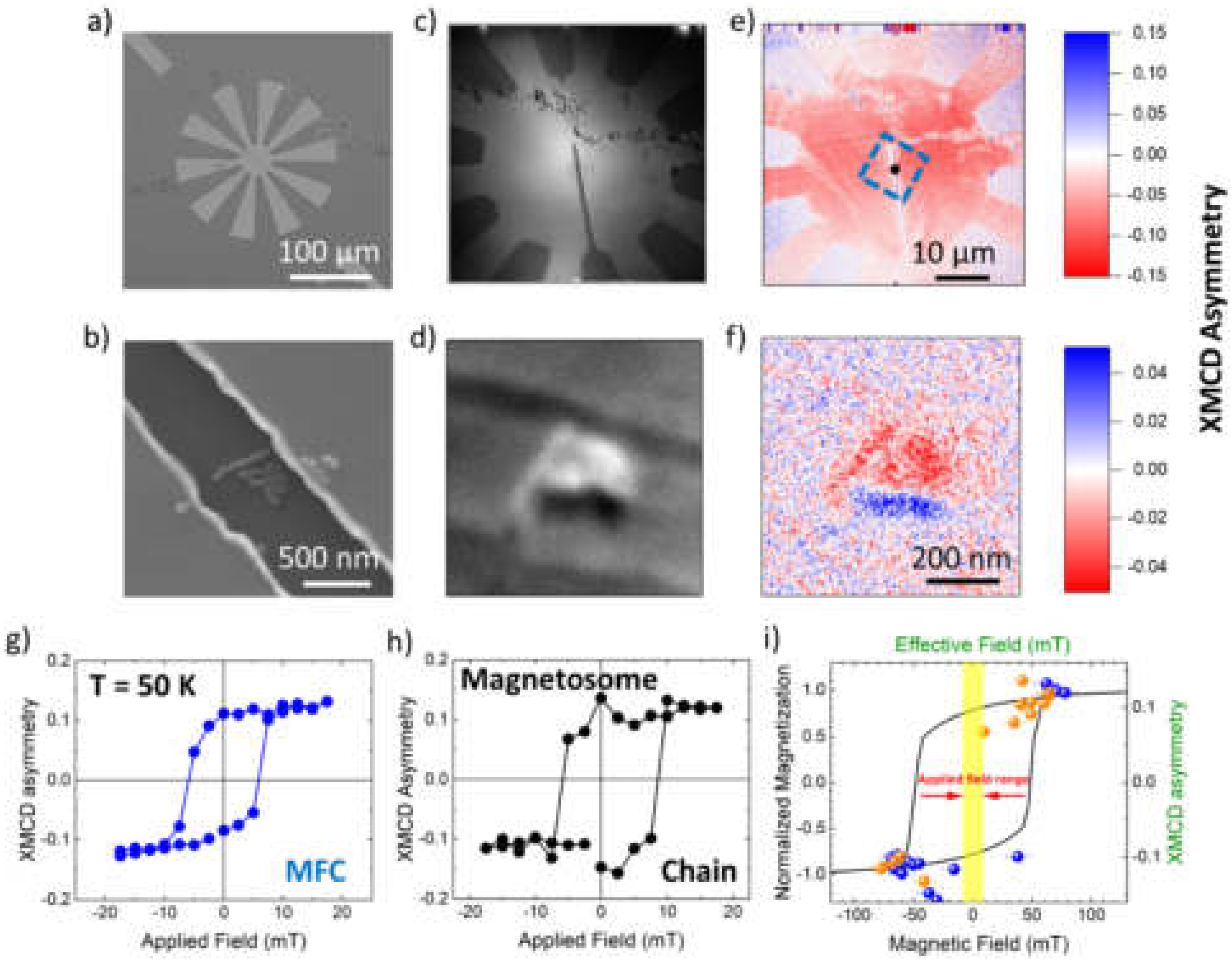

**Figure 2** Magnetosome chain and magnetic flux concentrator. SEM images showing a) the magnetic flux concentrator and b) the magnetosome chain positioned within its gap. The chain is oriented at ~20° relative to the magnetic field direction, which is perpendicular to the gap. c) and d) Room temperature XAS images of the concentrator (Co $L_3$-edge) and the chain (Fe $L_3$-edge), respectively. e) and f) corresponding XMCD images obtained at $\mu_0 H_{app}$ = 0 mT after magnetic saturation at negative fields. Blue dashed square and black dot in e) show the region used for the XMCD integration for the MFC and the location of the magnetosome chain, respectively. g) and h) XMCD magnetic hysteresis loops obtained at T = 50 K for the concentrator and the magnetosome chain, respectively. Vertical error bars are of the size of the symbols. i) Black curve: simulated loop for the magnetosome chain at T = 50 K considering a 20° misalignment between field and chain direction. The yellow area delimits the applied magnetic field range of the PEEM measurements.

This apparent discrepancy arises because the magnetosome chain does not experience the externally applied field directly, but the much larger effective magnetic field generated in the gap by the MFC. To account for this effect, the effective field is obtained by using equation 1, where the concentrator contribution ($\mu_0 H_{con}$) is determined from the spatial distribution of the XMCD signal measured for the MFC as a function of the applied field (see Supporting Information and Figure S1). The corresponding applied-to-effective field conversion curve is shown in Figure S3a.

The XMCD data for the chain as a function of the resulting effective field are shown in Figure 2i. Blue and orange dots correspond to XMCD data obtained along the increasing and decreasing field branches (after saturation), respectively. The fair agreement, in this representation, of the experimental data with the theoretical curve demonstrates that the presence of the MFC effectively concentrates the applied magnetic field, and enables accessing the complete space-resolved magnetic

,

hysteresis loop of the magnetosome chain within the limited field range available during field imaging. Indeed, effective fields close to ±100 mT can be achieved with applied fields an order of magnitude smaller (Figure S3a). This confirms that MFCs can extend the accessible magnetic field range, opening the door to studying systems that were previously inaccessible due to their high coercivity or saturation fields.

With this approach established, we next investigate the field-dependent magnetic domain structure of a giant magnetofossil[41], a system that has not yet been individually magnetically characterized owing to the challenges of isolating individual specimens and the high fields required for magnetic saturation[42].

**2.3 Application of on-chip MFC to high-field imaging of a giant magnetofossil**

While there is no clear consensus, the biogenic origin of giant magnetofossils, dating back at least 97 million years[41,43,44] ago, is supported, among others, by their widespread occurrence, their chemical purity, and crystallographic perfection[42]. More recently, insights into their 3D magnetic domain structure suggesting their use for magnetic navigation have added another piece of evidence for their biological origin from a magnetic perspective[45]. Beyond the work of Harrison et al. (Ref. [45]) there is no direct experimental imaging of their magnetic domain structure. Moreover, their magnetic domain characterization as a function of magnetic field is nonexistent, thus precluding confirmation of simulated magnetic hysteresis loops addressing relevant aspects such as the orientation of the easy- and hard-magnetic axis. In this context, magnetic imaging techniques represent an appealing possibility for the magnetic characterization of individual giant magnetofossils, provided that the available magnetic fields exceed their saturation fields, often being larger than ±100 mT[42].

A spearhead-shaped magnetite giant magnetofossil (Figure 3) with a length of approximately 2 µm and a base diameter of 1 µm was investigated by means of field-dependent XMCD-PEEM. To ensure sufficient flux concentration for this micro-scale specimen, the MFC was fabricated with the same lateral dimensions and the number of petals as in the previous case, but with an increased thickness of $t$ = 400 nm. This modification partially compensates for the expected reduction in the effective field associated with a wider nominal gap required to accommodate the fossil ($d_{gap}$ = 2 µm). From both, a reluctance-path viewpoint and micromagnetic considerations, increasing the MFC thickness enlarges the magnetic cross-section and magnetic volume of the device, thereby strengthening the stray-field contribution within the gap. The thickness dependence, together with the influence of other key geometrical parameters, is examined further in Section 2.4. During fabrication, cobalt deposition partially covered the tip of the fossil (Figure 3), which prevented its observation in the Fe $L_3$-edge XMCD-PEEM images.

,

The integrated XMCD signal as a function of $\mu_0 H_{app}$, for both concentrator and the fossil are depicted in Figure 3g and Figure 3h, respectively. The hysteresis loop obtained from the fossil (Figure 3h) differs significantly from those predicted by micromagnetic simulations (Figure 3i) considering most common crystallographic orientations along the fossil´s long axis[42], namely [113], [110], and [111], which predict negligible magnetic remanence and coercivity as well as a saturation field of approximately 100 mT. In similar way to the magnetosome chain, this apparent discrepancy results from the fact that the field experienced by the fossil is modified by the presence of the MFC. Indeed, both the MFC and the fossil exhibit similar loops with $\mu_0 H_c$ ~ 20 mT and $\mu_0 H_s$ beyond the 30 mT experimental limit.

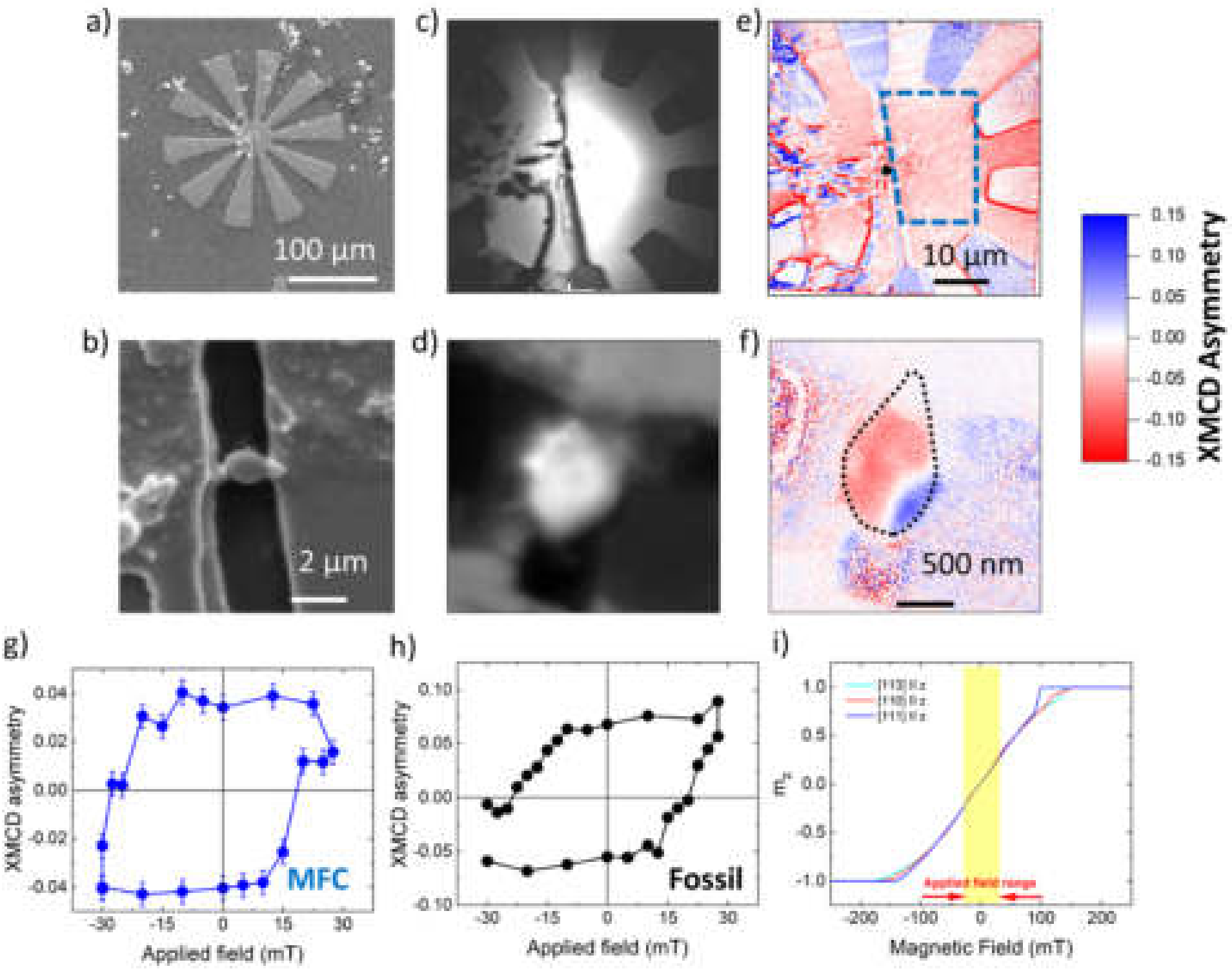

**Figure 3** Giant magnetofossil within a magnetic flux concentrator. SEM images showing a) the magnetic flux concentrator, and b) the fossil positioned within the 2 µm wide gap of the Co concentrator. The fossil is oriented parallel to the magnetic field direction, which is perpendicular to the gap. c) and d) XAS images of the concentrator (Co $L_3$-edge) and the fossil (Fe $L_3$-edge), respectively. Vertical error bars in h) are of the size of the symbols. e) and f) corresponding XMCD images obtained at $\mu_0 H_{app}$ = +15 mT. The contour of the giant magnetofossil is outlined by the dotted line. The covering of the fossil´s tip by Co during the fabrication of the MFC prevents its observation. Blue dashed right trapezoid and black dot in e) show the region used for the XMCD integration of the MFC and the location of the fossil, respectively. The area enclosed by the black dotted line in e) shows region for XMCD integration of the fossil. g) and h) XMCD magnetic hysteresis loops obtained for the concentrator and the fossil, respectively. Vertical error bars in h) are of the size of the symbols. i) Simulated loops assuming different crystallographic orientations along the long axis of the particle. The yellow zone delimits the magnetic field range of the PEEM without MFC.

,

Following the same procedure used above for the magnetosome chain, we convert the applied field into the effective field experienced by the fossil using the XMCD-derived magnetization of the concentrator (Supporting Information, Figure S3b). The resulting hysteresis loop is shown in the main panel of Figure 4. Notably, the maximum effective field reaches approximately 150 mT, about five times larger than the maximum field that can be delivered by the instrument during imaging without the use of MFCs (vertical light-yellow area in Figure 4j). Taking this field amplification into account brings the measured hysteresis into agreement with the micromagnetic simulations.

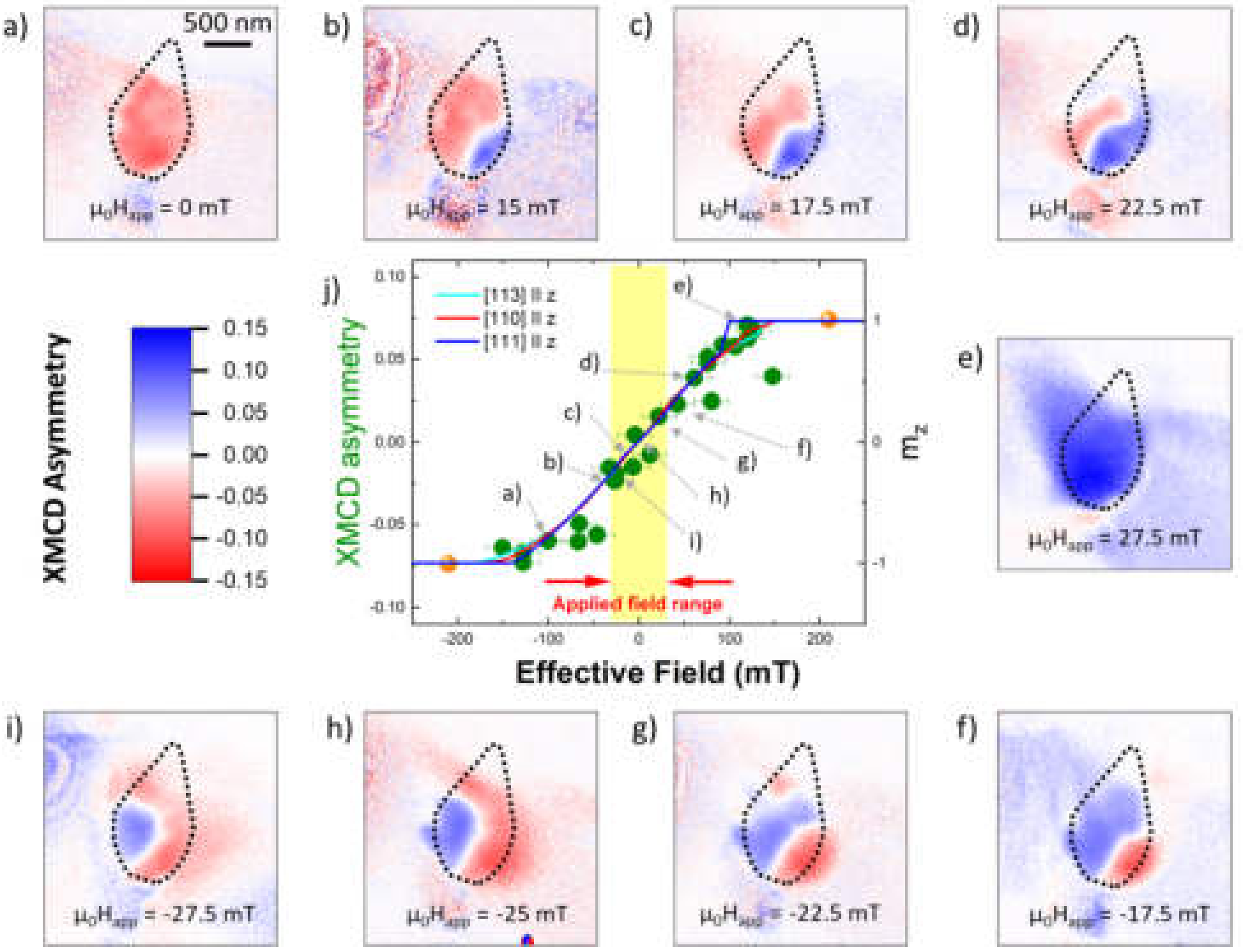

**Figure 4** Fe $L_3$-edge XMCD from giant magnetofossil vs effective field. Covering of the fossil´s tip by Co during the fabrication of the MFC prevents its observation. a)-i) spatially resolved XMCD images of the fossil (Fe $L_3$-edge) as a function of applied magnetic field, as indicated. The contour of the giant magnetofossil is outlined by the dotted line. j) XMCD hysteresis loops vs effective field obtained for the fossil. Gray arrows indicate points corresponding to XMCD images a)-i). The width of the light-yellow vertical bar signals the applied field range of the PEEM sample holder, see Figure 1. The maximum effective magnetic field during imaging with integrated MFC reaches ±150 mT. This is approximately 5 times larger than the maximum field available without MFCs. Orange points indicate effective field pulse values used to saturate the fossil at positive and negative fields (see Methods).

Despite the very good agreement between the effective field vs XMCD curve and the simulated loops (Figure 4j), the strong similarity among the simulated hysteresis curves for the [113], [110], and [111] crystallographic orientations along the fossil long axis precludes unambiguous identification. Moreover, although the agreement between experiment and simulations is good, it should be recalled that XMCD-PEEM is a surface sensitive technique, whereas the simulated loops depicted in Figures 3i

,

and 4j are averaged over the full volume of the particle. As shown in Figures S4 and S5, when the average magnetization is restricted to a near 5 nm thick surface region of the fossil, the simulated hysteresis strongly depends on azimuthal orientation. Consequently, a comparison between averaged surface-sensitive XMCD data and surface-restricted simulated curves becomes non-trivial due to this rotational degree of freedom.

This ambiguity, precluding a one-to-one identification between experiment and simulation, can instead be resolved by comparing the spatially resolved magnetization dynamics and reversal mechanism as a function of the applied field. In particular, the magnetic domain patterns imaged by XMCD-PEEM (Figure 4a-i) can be directly compared with those predicted by micromagnetic simulations (Figure 5, Figure S6 and Supplementary movies 1 to 12). In this context, the partial Co coverage of the tip of the fossil may slightly modify the coercive field but is not expected to significantly affect the reversal mechanism itself, see Supporting Information Figure S8.

Experimentally we observed that; starting from saturation at negative fields (Figure 4a), increasing the effective field leads to the nucleation of a new magnetic domain with magnetization component along the long axis of the fossil (*z*), antiparallel to the original domain (Figure 4b). This domain grows progressively as the effective field increases (Figure 4c-d), eventually reaching saturation (Figure 4e). A similar evolution is observed after saturating at positive effective fields and decreasing the field towards negative values (Figure 4f-i). The observation of two magnetic domains with opposite magnetization directions is compatible with a three-dimensional magnetic vortex ground state[45] (Figure S10), with its magnetic field evolution being dependent on the crystallographic orientation (Figures 5 and S4).

The experimentally observed magnetization reversal mechanism qualitatively agrees with simulations assuming a [113] crystallographic orientation along the long axis of the fossil (Figures 5, top panel). Moreover, this agreement also holds quantitatively when tracking the magnitude of the z-component of the magnetization during the reversal process. Figure S9 compares the field-dependent spatial integration of the absolute value of the z-component of the fossil's magnetization $|M_z|$ derived from XMCD measurements with that obtained from simulations. For comparison with the experiment the integration has been restricted to a near-surface region of 5 nm and the tip of the fossil has been excluded. For a [111] orientation, the simulations predict a decrease in the integrated $|M_z|$ value near zero field that is larger than experimentally observed for the integrated |XMCD| signal. Both [113] and [110] cases show similar trends, however the simulated magnetization maps for the [110] orientation (bottom panels of Figures 5 and S6) display extended regions of the fossil where the z-component of the magnetization is nearly zero. Experimentally, this would show up as the presence

,

of large areas with XMCD close to zero (whitish areas), which are not observed (Figure 4). Therefore, our experimental results are consistent with the spatially resolved magnetic domain structure and magnetization dynamics expected for a fossil with a [113] crystallographic orientation along its long axis. This conclusion is in agreement with recent findings obtained by selected-area electron diffraction and high-resolution transmission electron microscopy[42].

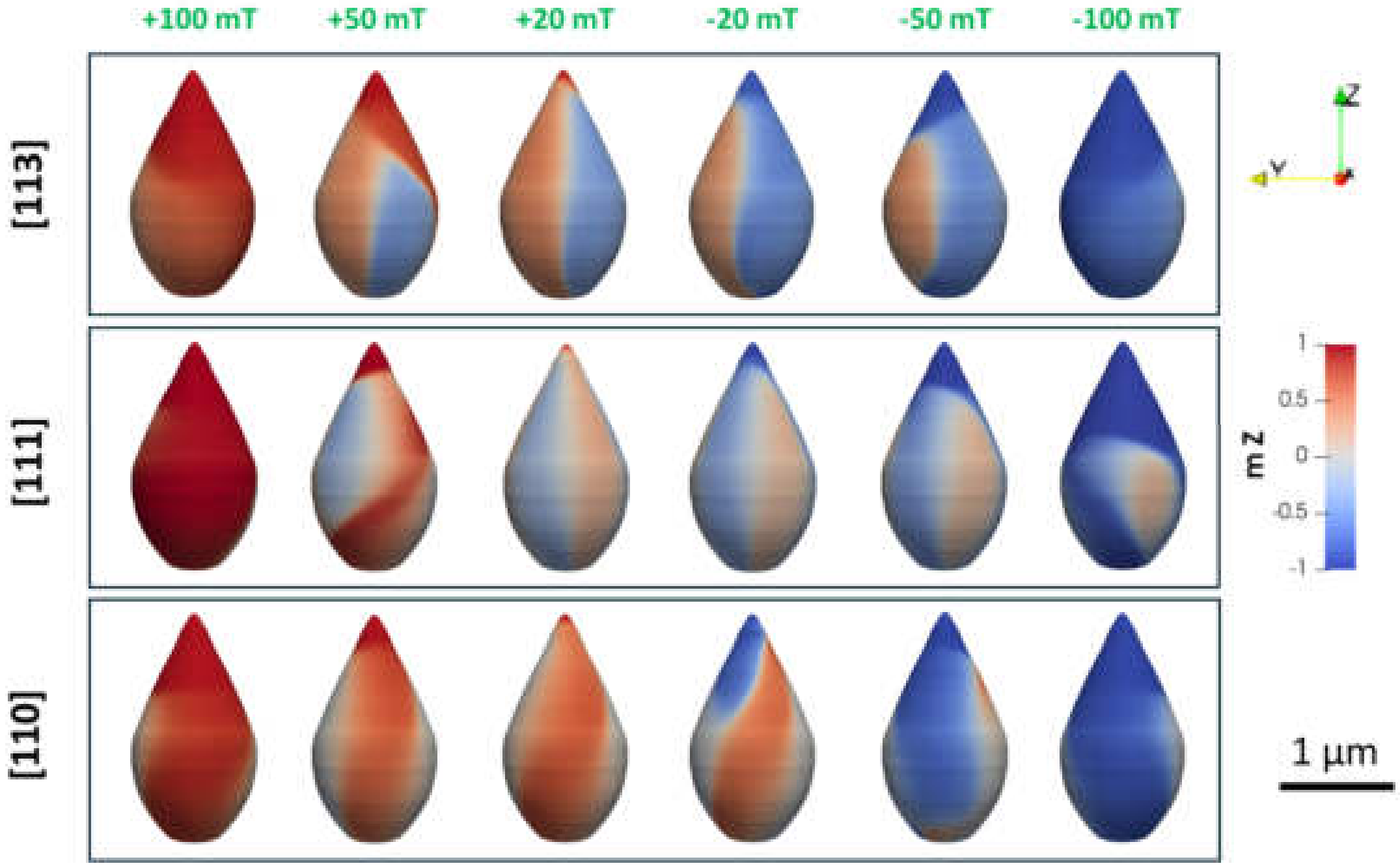

**Figure 5** Micromagnetic simulations of the imaged spearhead giant magnetofossil grain considering different crystallographic orientations. View along +x direction of the z-component of the magnetization of the giant magnetofossil as a function of a magnetic field applied along its long axis (z) after magnetic saturation at +300 mT. Top: [113] crystallographic orientation along z, middle: [111] crystallographic orientation along z, and bottom: [110] crystallographic orientation along z. Magnetic domain evolution with field for [113] agrees with experimental observations, see Figure 4.

### 2.4. Key parameters governing the magnetic flux concentrator performance

The performance of an MFC depends primarily on a few geometrical and material parameters that determine how efficiently magnetic flux is guided and concentrated into the gap. While the qualitative behavior is described in Eq. (1), selecting an appropriate concentrator for a given magnetic system requires understanding how its most influential design parameters affect the attainable effective field. Here we briefly summarize the main trends identified from micromagnetic and magnetostatic simulations; full details and extended parameter sweeps are provided in the Supporting Information.

,

A first determinant of concentration efficiency is the thickness of the ferromagnetic film (Figure 6a). Increasing the thickness enhances both the magnetic volume and the cross-section available for the flux-guiding path, leading to a stronger stray-field contribution within the gap. Thicker devices also exhibit a more gradual magnetization-reversal behavior, which is advantageous for achieving stable and reproducible effective fields. This parameter becomes particularly relevant when the gap must be widened to accommodate larger magnetic structures. The gap width directly controls the achievable effective field: smaller gaps provide stronger concentration, whereas larger gaps reduce the local field amplitude. In practice, the gap must be chosen to match the footprint of the sample while remaining as small as feasible (Figure 6b). The material properties of the concentrator influence both the maximum effective field and the switching behavior. In particular, higher saturation magnetization leads to stronger field concentration, while the exchange stiffness affects coercivity and domain-wall stability. Among the materials tested, Cobalt provides the highest effective fields (Figure 6c).

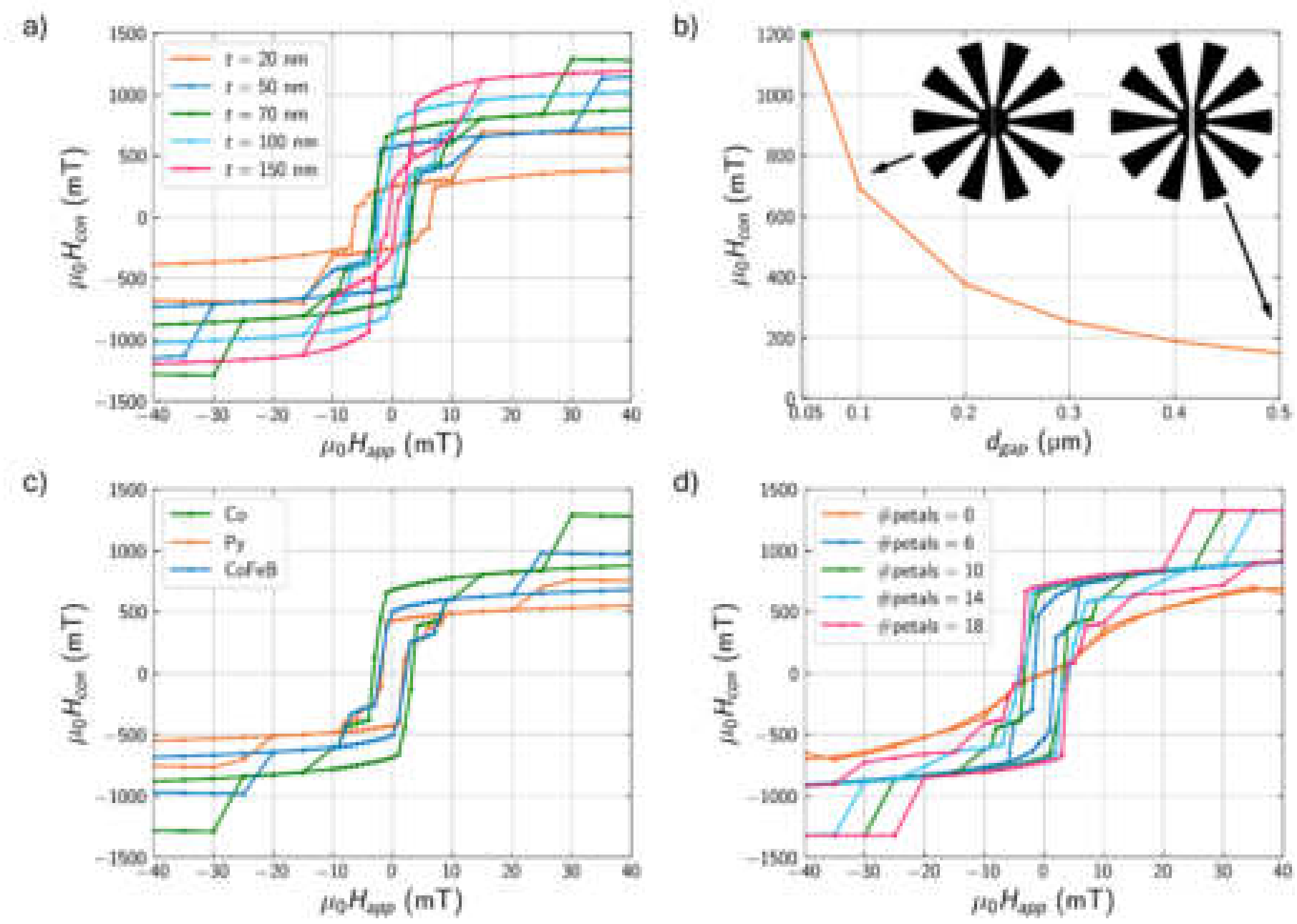

**Figure 6** Micromagnetic simulations of the field created by the MFC ($H_{con}$) at its gap center as function of applied field ($H_{app}$) for different geometric and material parameters, varying; a) thickness, b) gap width, c) material, and d) number of petals. Other parameters are fixed unless otherwise stated: $R_O = 5R_i = 10$, $t = 70$ nm and $d_{gap}$ = 50 nm, cobalt. Field applied across the gap.

The petal geometry plays a key role in the flux-guiding behavior of the concentrator (Figure 6d). Simulations show that MFCs without petals exhibit a nearly linear, non-hysteretic response and become effective only at comparatively high applied fields, where the ferromagnetic film approaches saturation, and the field at the gap is dominated by the local stray field. Introducing petals, however, enables magnetic-flux concentration already at lower applied fields and modifies the reversal process: increasing the number of petals tends to enhance the anisotropic flux guidance and produces larger

,

remanence, slightly higher coercivity, and a greater number of intermediate magnetization states during switching. Conversely, reducing the number of petals generally leads to smoother reversal and lower remanence, though at the cost of reduced low-field concentration efficiency. Variations in petal length mainly affect the width of the switching-field interval, with only modest influence on remanence or coercivity for sufficiently long petals, see Supporting Information. Overall, the petal structure governs the balance between strong low-field concentration and the complexity of the magnetization-reversal process.

To summarize, an efficient MFC should combine sufficient magnetic volume (increase thickness and outer radius) to generate a high demagnetization field in the MFC gap, a petal geometry that provides strong anisotropic flux guidance with smooth and stable magnetization switching, the smallest gap size compatible with the specimen, and a material with high saturation magnetization. These elements ensure a high effective field at saturation, low coercivity, and a low applied field required for switching. Depending on the size of the magnetic structure under investigation and the type of field-dependent characterization to be performed, these parameters can be adjusted to favor stronger concentration, smoother reversal, or lower operating fields, providing a practical and flexible framework for tailoring MFC designs to a broad range of magnetic systems.

## 3. Conclusions

Magnetic field concentration based on transformation optics has previously been used to enhance the efficiency of magnetic sensors. Here, we demonstrate that this concept can be implemented as a non-invasive approach to extend the accessible magnetic field range in magnetic microscopy techniques. This is implemented using flower-shaped MFCs, which provide a practical realization of the required anisotropic permeability, while not being the only possible geometry.

Scaling down the magnetic concentrator to micrometer and sub-micrometer scale necessarily imposes constraints on the design of the MFC and, consequently, on its final performance. Factors such as sample dimensions and specific requirements of the experimental technique employed may limit, for example, the gap width or thickness of the MFC. While this inherently restricts the magnetic gain and makes it sample and/or experiment dependent, it nonetheless ensures an improvement over the current state of the art.

These design constraints, together with micromagnetic details and imperfections introduced during the fabrication of the MFC and/or related to sample preparation, inevitably affect the precise determination of the effective field at the sample position, as such effects cannot be fully captured—

,

even when spatially resolved magnetic information of the MFC itself is available, as in the present case. Nevertheless, the resulting calibration curves, whether obtained from independent measurements or directly during the acquisition of the sample's magnetic loop, remain sufficiently close to the true field response to extract meaningful magnetic information. Moreover, in many magnetic-sensitive microscopy studies, the key interest lies in qualitatively resolving magnetic textures and how do they evolve with field: the nucleation and annihilation of domains, the stability of topological structures, the comparison of switching behavior across different elements or layers, and the identification of field-induced transitions. For these purposes, the gain obtained by significantly extending the available magnetic field range is worth even at expense of inaccurate quantitative estimation of the field values. This is precisely the case for the fossil data we have presented, where the correlation between the experimentally observed magnetization reversal and micromagnetic simulations enables identification of the crystallographic orientation, even without exact knowledge of the effective field.

Within this context we show that engineering the magnetic properties and geometry of ferromagnetic metamaterial-inspired, flower-like magnetic flux concentrators, it is possible to expand the accessible field range in PEEM-based magnetic imaging. Our results demonstrate that sample-integrated MFC elements can extend the available magnetic field range well beyond its current limit. Moreover, the possibility of imaging at large effective magnetic fields, resulting from the concentration of significantly smaller applied fields, reduces not only sample heating but also minimizes electron deflection, thereby improving spatial resolution as shown within the Supporting Information.

More broadly, this strategy holds promise for a range of magnetic imaging techniques where external magnetic field strength is limited by instrumental constraints. For instance, MFCs could enable stronger local fields in electron-based microscopes such as Lorentz transmission electron microscopy and scanning electron microscopy with polarization analysis, by minimizing interactions between the magnetic field and electron optics. As in PEEM, this would be made possible by the strong spatial confinement of the magnetic field generated by the MFC, which decays within approximately 1 µm from the sample surface. Such a near-surface confinement could also benefit techniques such as magnetic force microscopy and nitrogen-vacancy (NV) center microscopy, where the probe response can be influenced by applied fields[20]. Furthermore, techniques with geometric or spatial limitations, such as X-ray transmission microscopy, X-ray ptychography, and X-ray laminography, could benefit from the microscale dimensions of MFCs and their direct sample integration.

,

## Acknowledgements

We thank the Helmholtz-Zentrum Berlin für Materialien und Energie for the allocation of synchrotron radiation beamtime. AP and LlB acknowledge financial support from MCI and MCIN/ AEI /10.13039/501100011033/ through CHIST-ERA PCI2021-122028-2A co-financed by the European Union Next Generation EU/PRTR, the “Severo Ochoa” Programme for Centres of Excellence CEX2023-001263-S, HTSUPERFUN PID2021-124680OB-I00, HTS-4ICT PID2024-156025OB-I00 and ETMOS (PID2024-155387OB-I00), co-financed by ERDF A way of making Europe. The Spanish Nanolito networking project (RED2022-134096-T). AP and AVS thank the support from the European COST Action SUPERQUMAP (CA 21144). AB acknowledges support from MICIN Predoctoral Fellowship (PRE2019-09781). The authors acknowledge the Scientific Services at ICMAB and the UAB PhD program in Materials Science. AVS acknowledges financial support from Fonds de la Recherche Scientifique - FNRS under the programs PDR T.0204.21 and EraNet-CHISTERA. AVS and EF acknowledge financial support by the FWO and F.R.S.-FNRS under the Excellence of Science (EOS) project O.0028.22. LM thanks the Spanish State Research Agency (AEI) / 10.13039/501100011033 and by the Ministry of Science and Innovation under project MCINN-24-PID2023-150968OA-I00 and the Research Agency of the Principality of Asturias(SEKUENS) under projects SEK-25-GRU-GIC-24-113. RJH acknowledges funding from the 714 Royal Society Newton Advanced Fellowship (grant NAF\R1\201096). L.C. thanks funding support from the National Natural Science Foundation of China (grant 42330204). MLG acknowledges support by the Spanish MCIN/AEI/10.13039/501100011033 under project PID2023-146448OB-21, and the Basque Government under project IT-1479-22. DV gratefully acknowledges the grant PRE2021-099247 funded by MCIN/AEI/10.13039/501100011033 and by “ESF Investing in your future”. The work of TP and BV was supported in part by the Fonds de la Recherche Scientifique - FNRS under Grant n°J.0232.26

## Author contributions

SV, AP and AVS conceive the experiment. LC and RH provided giant magnetofossils samples and located them by means of SEM. LM, DV, and AG isolated the magnetosomes from bacteria and prepared the magnetosome chain. LM and MLF-G prepared and characterized magnetosome samples. AB, LlB, and AP prepared the magnetic flux concentrators. RJH did micromagnetic simulations of the fossils. EF, TP, BV and AVS performed micromagnetic simulations of the MFC. IO simulated the magnetic hysteresis loop of the magnetosome chain. SV and RA performed the X-PEEM experiments. SV analyzed the synchrotron data. All authors discussed the results. SV and AP wrote the manuscript with input from all authors.

## Methods

,

*a) Giant magnetofossil sample preparation:*

The investigated giant magnetofossil sample originates from the North Atlantic Integrated Ocean Drilling Programs (IODP) Site U1403 (~4374 m paleodepth)[34]. Magnetic mineral extracts were obtained by magnetic separation[43]. Isolated fossils were localized by means of scanning electron microscopy (SEM).

*b) Magnetosome sample preparation:*

*Magnetospirillum gryphiswaldense* MSR-1 (DMSZ 6631) was cultivated at 28 °C without shaking in flask standard medium supplemented with 100 µM Fe(III) citrate[46]. After 48 h-incubation, when bacteria present well-formed magnetosome chains, the cells were harvested by centrifugation at 8000 G for 15 minutes and suspended in 20 mM HEPES–4 mM EDTA (pH 7.4). Magnetosomes were isolated according to the protocol described by Grünberg et *al.*[47] with minor modifications. Cellular disruption was performed using a French press at 1250 psig and sonication at 40 W for 30 minutes (cycles of 15"/5" ON/OFF). The magnetosomes were subsequently separated from cell debris by magnetic separation and rinsed 10 times with 10 mM Hepes–200 mM NaCl (pH = 7.4), with further sonication (40 W, 15 minutes, cycles of 15"/5" ON/OFF) in between rounds of magnetic separation. The final magnetosome suspension was adjusted to a concentration of, 30 µg $mL^{-1}$ in ultra-pure water. 5 µL of the magnetosome suspension was deposited onto a Si substrate under an externally applied magnetic field of 0.4 T to promote chain alignment. To ensure a homogeneous distribution of magnetosomes and to minimize surface tension effects, infrared radiation was used during the drying process.

*c) Magnetic flux concentrator:*

Once the magnetic structures were deposited onto the Si substrates, the samples were examined and characterized with an SEM and an optical microscope to localize the targeted structures. Subsequently, the MFCs were patterned and aligned to the structures by photolithography using a bilayer resist (Shipley 1813 as the upper layer and polydimethylglutarimide (PMGI) as the bottom layer). Co was then deposited by DC magnetron sputtering at room temperature with a pressure of $6.10^{-3}$ mbar in an argon atmosphere and a power of 10 W. The lift-off process was carried out in an acetone bath, assisted by pipetting, to remove the upper resist and the undesired Co. Finally, the samples were immersed for 1 min in propylene glycol monomethyl ether acetate to remove the bottom layer and rinsed in distilled water to eliminate residual solvents.

*d) XMCD-PEEM experimental setup*

The silicon substrates with integrated MFCs and the sample to be imaged (magnetosome chain or giant magnetofossil) were mounted on a sample holder capable of applying magnetic fields, see Figure

,

1a. Correction of deflected electron trajectories by PEEM electron optics allows in-field imaging up to ca. 30 mT (Ref. [19]).

The gap of the MFC was aligned parallel to and positioned within the magnetic gap of the sample holder. The incoming radiation impinged on the sample at 16° grazing incidence, with its propagation direction being parallel to the gap, and therefore parallel to both the applied and effective magnetic field directions. In this configuration the X-ray magnetic circular dichroism is proportional to the magnetization component across the gap direction.

XMCD-PEEM experiments were done at the PEEM station at the UE49/PGMa beam line of the synchrotron radiation source BESSY II of the Helmholtz-Zentrum Berlin[48]. The incoming photon energy was set to 777.8 eV (Co $L_3$-edge) for imaging the MFCs and to 707.6 eV (Fe $L_3$-edge) for measuring the magnetosome chain and giant magnetofossil magnetite samples. Field-dependent imaging was done at room temperature for the giant magnetofossil sample and at T = 50 K for the magnetosome chain.

e) *Field-dependent XMCD measurements*

A magnetic field pulse of -100 mT was applied before data collection for magnetic hysteresis loops started at negative fields. After reaching the experimental limit of +30 mT a magnetic field pulse of +100 mT was applied before collecting data at decreasing fields. The presence of the MFC generates larger effective fields ensuring magnetic saturation of the magnetite samples. Applied fields pulses of ± 100 mT correspond to effective field pulses of approximately 210 mT (orange dots in Figure 4j), a factor two larger than currently available.

At each magnetic field step, 80 images (3 s integration each) for the magnetite samples and 40 for the MFCs were acquired for incoming circular polarized radiation with left ($\sigma^-$) and with right ($\sigma^-$) helicity. Each image was normalized to a bright field image and drift corrected before its averaging. The magnetically sensitive XMCD asymmetry images were obtained as $(\sigma^- - \sigma^+)/(\sigma^- + \sigma^+)$. XAS images depicted in Figures 2 and 3 are obtained by computing XAS = $\sigma^- + \sigma^+$.

*f) Simulation of the magnetic hysteresis loop for the magnetosome chain*

Magnetic hysteresis loops were obtained using a dynamical approach[39] where magnetosomes along the chain are considered as a collection of independent single domain particles. These particles are large enough to be thermally stable, ensuring that the magnetization is firmly anchored at the minimum energy states. The equilibrium orientation of each magnetic moment was obtained by minimizing the single dipole energy density[40]. At 50 K, below the Verwey transition temperature ($T_V$), the magnetocrystalline anisotropy changes from cubic to essentially uniaxial oriented to the square faces of magnetosomes, aligned along the ⟨100⟩ directions of the original cubic spinel lattice. In this

,

regime, the effective anisotropy is thus purely uniaxial ($K_{uni} = 16\ kJ/m^3$), arising from the competition between the magnetocrystalline uniaxial anisotropy, shape anisotropy and dipolar interactions with neighboring magnetosomes. The experimentally observed misalignment of ca. 20° between the field and the <111> chain directions was considered, i.e. t the hysteresis loop depicted in Figure 2i resulted from the averaging of 18 cycles (up to 80kA/m) in which the field directions lie on a conical surface with an opening angle of 20°.

*g) Micromagnetic simulations of the magnetic flux concentrator*

Micromagnetic simulations were performed using the open-source software MuMax3 (Ref. [49]), which employs the finite-difference method to compute the time- and space-dependent magnetization evolution by solving the Landau–Lifshitz–Gilbert (LLG) equation. The investigated magnetic structure is assumed to be polycrystalline Co, characterized by an exchange constant $A_{ex}$=30 pJ/m and the saturation magnetization $M_s$=1120 kA/m. The discretization cell size was set to 10 nm in each direction. The use of a cell size slightly larger than the exchange length in Co (6 nm) has been addressed with smaller devices, showing that it introduces only negligible errors in the stray field concentrated in the gap, and is therefore suitable to capture the concentration gain of our device. Moreover, since micromagnetic simulations are not well suited for structures in the order of hundreds of micrometers (due to limited GPU memory), the in-plane dimensions of the simulated structure were scaled down by a factor of 10 compared to the experimental devices. Care was taken to preserve the aspect ratio (to minimize differences in the demagnetization factor between experiment and simulation) and to choose the sample thickness such that the same type of domain walls is obtained. Additionally, to avoid unrealistic results caused by numerical perfect symmetry of the device, a uniform deviation of 1° was introduced in the direction of the applied field.

As pointed out above, in order to make the simulations computationally tractable, the lateral dimensions of the MFCs were scaled down by a factor of ten. At these reduced scales, the magnetic reversal process involves a limited number of domains, leading to sudden jumps in the magnetization as a function of the applied field. In contrast to that, the experimentally realized MFCs are significantly larger, allowing for a larger number of domains and a more gradual and spatially averaged magnetization reversal, which smooths out such discontinuities.

*i) Micromagnetic simulations of the giant magnetofossil*

Finite-element micromagnetic simulations of the giant magnetofossil were performed using MERRILL[50]. Micromagnetic parameters for magnetite at room temperature were saturation magnetization, $M_s$ = 4.80768 x $10^5$ A/m, first cubic magnetocrystalline anisotropy constant, $K_1$ = -1.32658 x $10^4$ J/m$^3$, and exchange constant, $A_{ex}$ = 1.33487 x $10^{-11}$ J/m. A 3D model of the particle was

,

created by tracing the two-dimensional outline of the particle from scanning electron microscopy images, measuring the radius of the particle as a function of position along its length, and then building a 3D model from a series of stacked circular frustrums with height 20 nm each. The particle was 2.07 µm long and 1.14 µm wide at its widest point. Tetrahedral meshes were generated from the 3D models using a mesh resolution of 20 nm using Coreform Cubit software. Simulations were performed with either the [110], [111] or [113] cubic crystallographic directions aligned with the length of the particle (111 is the magnetocrystalline easy axis for magnetite). Simulations of the upper branch of the hysteresis loop were performed by conjugate gradient energy minimization for fields between +300 mT and -300 mT in steps of -10 mT applied along the length of the particle ($z$). Simulation results were visualized using Paraview[51].

*j) Determination of the effective field created by the MFCs*

As described within the text, the effective field is the sum of the applied field and the field created by the magnetic flux concentrator at the sample position. The latter has been computed from the spatially resolved XMCD data of the MFC. As shown within the Supporting Information, the z-component of the concentrated field can be derived from the z-component of the magnetization, which is proportional to the XMCD.

The error in the determination of the effective field thus arises from the error in the XMCD. We consider a uncertainty in the XMCD asymmetry determination of ±0.005. This error corresponds to 12.5 % of the XMCD value at saturation for the MFC with the fossil and approximately 4 % for the MFC used for the magnetosome chain. Given the relation between the z-component of the magnetization, the XMCD and the field created by the MFC, this translates to an error determination of the field created by the concentrator of 12.5% and 4% of its value at saturation (~ 110 mT and 80 mT, respectively) which means uncertainty for $H_{con}$ and so for $H_{eff}$ of ± 14 mT and ±5 mT.

For the MFC corresponding to the fossil and the magnetosome chain, this error is incremented by ±5 mT and ±4 mT, respectively, due to the presence of foreign material close to the sample position which might locally modify the effective field via dipolar interactions, see Supporting Information.

,

,

,